\documentclass[pra,twocolumn,superscriptaddress,showpacs]{revtex4-1}
\usepackage{amsmath,amssymb} 
\usepackage{bm}
\usepackage{hyperref}
\usepackage{subfigure}
\usepackage{graphicx}
\usepackage{color}

\begin{document}

\title{Observations of $\lambda/4$ structure in a low-loss radiofrequency-dressed optical lattice}

\author{N.~Lundblad}
\email{nlundbla@bates.edu}
\noaffiliation
\author{S.~Ansari}
\author{Y.~Guo}
\author{E.~Moan}
\noaffiliation
\affiliation{Department of Physics and Astronomy, Bates College, Lewiston, ME 04240, USA}

\date{\today}
\begin{abstract}

We load a Bose-Einstein condensate into a one-dimensional (1D) optical lattice altered through the use of radiofrequency (rf) dressing.   The rf resonantly couples the three levels of the $^{87}$Rb $F=1$ manifold and combines with a spin-dependent ``bare" optical lattice to result in adiabatic potentials of variable shape, depth, and spatial frequency content.  We choose dressing parameters such that the altered lattice is stable over lifetimes exceeding tens of ms at higher depths than in previous work.   We observe significant differences between the BEC momentum distributions of the dressed lattice as compared to the bare lattice, and find general agreement with a 1D band structure calculation informed by the dressing parameters.   Previous work using such lattices was limited by very shallow dressed lattices and strong Landau-Zener tunnelling loss between adiabatic potentials, equivalent to failure of the adiabatic criterion.   In this work we operate with significantly stronger rf coupling (increasing the avoided-crossing gap between adiabatic potentials), observing dressed lifetimes of interest for optical lattice-based analogue solid-state physics.

\end{abstract}
\maketitle

The optical lattice is a versatile  tool for  trapping and control of neutral atoms and for studying both single-particle and many-body quantum physics.   It has proven useful to  optical atomic clock development~\cite{Derevianko:2011p3800}, to the development of quantum-computing proposals~\cite{Nelson:2007lr,Lundblad:2009gz,Lee:2013ko}, and to solid-state analogues and quantum simulations~\cite{Bloch:2012jy,RevModPhys.80.885}, including the ability to resolve these systems at the single-atom level~\cite{Gemelke:2009p1525,Endres:2011it,Simon:2011p3794}.      While early work focused on simple lattices of $\lambda/2$ periodicity (where $\lambda$ is the wavelength of the lattice laser), including square (2D) and cubic (3D) lattices, more complex periodic potentials were sought out in order to enhance existing lattice-physics experiments and to explore less well-understood many-body physics.     Recently, new lattice geometries (including the triangular, honeycomb, kagome, double-well, and checkerboard lattice) have been explored using  various techniques, including the use of dual commensurate lattice lasers~\cite{Windpassinger:2013is} as well as through holography~\cite{Klinger:2010p2742}.     Additionally, lattice substructure in 1D has  been generated using Raman transitions~\cite{salger:011605,ritt:063622,PhysRevLett.76.4689,Miles:2013hj}.

Taking a wholly different approach, other work~\cite{Lundblad:2008kb,Yi:2008hy, 2008PhRvA..78e1602S} introduced a method of altering lattice geometry and topology based on the notion of radiofrequency (rf) dressing of spin-dependent lattice potentials~\footnote{Microwave dressing of spin-dependent lattices has been explored in the context of spin-dependent transport as described in Ref.~\cite{Mischuck:2010gg}}.  The theoretical work~\cite{Yi:2008hy,2008PhRvA..78e1602S} aimed at the exploitation of the resulting adiabatic potentials' faster tunnelling timescales and higher interaction energies, as well as the associated higher temperature scale.    The experimental work~\cite{Lundblad:2008kb} showed that it was possible to generate 2D dressed lattices that in principle had tailored subwavelength structure, in particular pointing the way to toroidal single-site wavefunctions~\footnote{Toroidal wave functions have also been proposed in the context of a four-level tripod-type system~\cite{Ivanov:2014up}}.     This work was limited, however, to dressed lattices of rather small depth, preventing the realization of tight-binding lattice wavefunctions with lifetimes appropriate to the solid-state-analogue physics goals.     Attempts to push toward  deeper dressed lattices were blocked by nonadiabatic (Landau-Zener) losses associated with insufficient rf coupling strength and lattice-laser power.    In this Article we present observations of a 1D dressed optical lattice system in a regime where interesting structure (finer than the usual $\lambda$/2 scale) appears such that Landau-Zener losses are limited and the dressed lattice  begins to be of useful depth.    We demonstrate that the dressing procedure alters the momentum distribution of a BEC loaded into the lattice in a way that generally agrees with a band-structure calculation informed by the dressing parameters, and that the adiabatic potentials are stable over $\gtrsim$ 30 ms timescales (of similar order to tunnelling and interaction timescales) given sufficient dressing (rf coupling) strength.

\begin{figure}[t!]
\centering
  \includegraphics[width=\columnwidth]{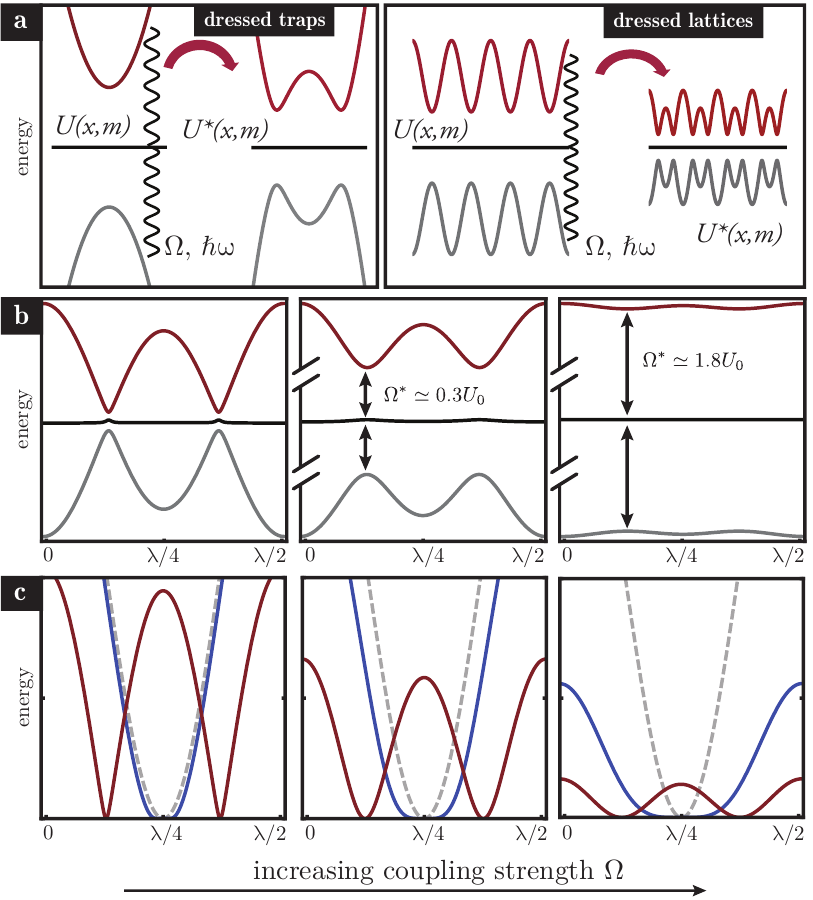}\caption{
  {\bf (a)} The general scheme for radiofrequency (rf) dressing of a symmetric three-level system, in the case of a quadratic magnetic-trap potential (at left) and a periodic (lattice) potential (at right), both at rf detunings near resonance.     In both cases the bare spin-dependent potentials $U(x,m)$ become (in the dressed picture, with rf frequency $\omega$ and rf coupling strength $\Omega$) the adiabatic potentials $U^*(x,m)$.   In the case of the dressed lattice note the appearance of $\lambda/4$ structure in the adiabatic potentials.
  {\bf (b)} Left-to-right: calculated lattice adiabatic potentials for $\Omega$ = $U_0$/16, $U_0$/2.5, and 2.5$U_0$, where $U_0$ is the bare lattice depth.   The maximum depth of a dressed lattice (for $\Omega\rightarrow 0$) is $U_0/2$, i.e., depth decreases with increased $\Omega$.   The gap between adiabatic potentials, relevant for nonadiabatic transitions, increases with coupling strength as $\Omega^*\simeq\Omega/\sqrt{2}$.    {\bf (c)} The dependence of the uppermost adiabatic potential on the rf detuning $\delta$ for the same three values of $\Omega$ as in (b):  the bare potential (dashed),  the slightly altered lattice $\delta\simeq+U_0/8$  (blue), and a $\lambda/4$ lattice  $\delta \simeq+U_0/2$ (red).    }    
\label{setup}
\end{figure}

Radiofrequency dressing (and the notion of adiabatic potentials), as illustrated in Fig.~\ref{setup}, is familiar in the context of magnetic traps, where ``rf-knife" evaporative cooling can be viewed as the atomic traversal of the lower adiabatic potential of a dressed spin system~\cite{mitevap}.   Similarly, bubble- or shell-like potential surfaces can be created using the upper adiabatic potential~\cite{white:023616,PhysRevLett.86.1195,Merloti:2013ft} with possible additional time-averaging~\cite{Lesanovsky:2007kw,Gildemeister:2010gx}.  Coherent splitting of an atom-chip BEC for interferometry can also be achieved using rf dressing~\cite{Hofferberth:2006p749,lesanovsky:033619}.    We consider a BEC held in a  1D optical potential of the form
\begin{equation} U(x,m_F) = -m_F U_0 \cos^2(kx),
\end{equation}where $m_F\in\{0,\pm 1\}$ is the total angular momentum projection, $\hbar U_0$ is the lattice depth, and $k=2\pi/\lambda$ is the wavevector of the lattice laser at wavelength $\lambda$. A bias magnetic field $B_0$ is applied along $x$, resulting in linear Zeeman splitting given by $\hbar \omega_{-1,0}$ and $\hbar \omega_{0,+1}$.      In order to create the desired adiabatic potentials we couple the three $m_F$ levels with near-resonant rf radiation.    An rf magnetic field of amplitude $B_{\rm rf}$ and frequency $\omega$ is applied perpendicular to $B_0$.      If we make a Born-Oppenheimer-type approximation (ignoring residual kinetic energy) and take the usual rotating-wave approximation, the rotating-frame Hamiltonian $\mathcal{H}$ for the system is
\begin{equation}
\mathcal{H}({\bf r})=\left(\begin{array}{ccc}U(x,-1)-\delta & \Omega/2 & 0 \\ \Omega/2 & U(x,0) & \Omega/2 \\0 & \Omega/2 & U(x,+1)+\delta+\delta'\end{array}\right)
\end{equation}
where $\delta =\omega-\omega_{-1,0}$ is the rf detuning, $\hbar\Omega = \mu_B B_{\rm rf}/2$ is the standard coupling-strength matrix element  and $\hbar\delta'$ is a small deviation from the linear Zeeman regime (much smaller than $\Omega$ in our experiments).     Diagonalizing this Hamiltonian yields the so-called adiabatic eigenstates: a spatially varying superposition of the bare spin eigenstates.   Figure 1 depicts several illustrative cases of the resulting adiabatic potentials for different $\delta$ and $\Omega$.     Notably, we see that for $\delta\simeq\delta_0= +U_0/2$ the periodicity of the lattice is halved.    In order to observe atoms in such a lattice, we proceed as follows: generate a spin-polarized BEC, load it into the bare lattice, apply the lattice dressing procedure (discussed below), and image the momentum distribution of the system through ballistic expansion.

Our BEC apparatus is a hybrid  machine combining a magnetic trap and a single-beam optical dipole trap, largely following the design of Ref.~\cite{Lin:2009p1534}. A Zeeman slower delivers  $^{87}$Rb atoms to a conventional six-beam magneto-optical trap (MOT); the trapped sample (in the weak-field-seeking   5$^2 S_{1/2}$ $|F=1,m_F=-1\rangle$ state) is then transferred into a colocated magnetic quadrupole  trap and subject to rf evaporation.      The magnetic field zero of the quadrupole trap results in spin-flip ``Majorana" loss as the sample becomes colder, a problem that in this case is solved via transfer to a single-beam optical dipole trap aligned slightly below the field zero.   The transfer process notably provides an increase in phase-space density due to the drastic change in trap geometry, the statistical mechanics of which is detailed elsewhere~\cite{Lin:2009p1534}.    Further evaporation occurs through forced reduction of the dipole potential; in our apparatus BEC typically appears at a critical temperature $T_c\simeq 200$ nK, and condensates of $>$90\% purity with $\sim$$10^5$  atoms are regularly produced with an overall experimental cycle time of 30 s.

We load the BEC into the ground band of a 1D optical lattice with laser power (lattice depth) increasing exponentially over 300 $\mu$s.   This timescale was chosen to be  adiabatic with respect to vibrational excitation in the lattice~\cite{PhysRevA.67.051603}.   The circularly-polarized lattice beam is generated by a Ti:Sapphire laser held near 790.06 nm, the tune-out wavelength between the D1 and D2 lines of $^{87}$Rb where the light shifts for $m_F =\pm 1$ are opposite in sign and equal in magnitude, and the light shift for $m_F=0$ is absent, as presented in the lattice potential of Eq.~1.    The bare lattice is held for a few ms to allow for stabilization of the bias field,
and in the case of deep bare lattices, dephasing of the individual lattice sites.  The dressing process is initiated by switching on the rf coupling $\Omega$ (provided by a resonant loop antenna fed by a 150W broadband amplifier) at a fixed frequency $\omega$, with a large detuning $\delta$ set by the magnetic field $B_0$.    At this point the adiabatic potentials are indistinguishable in shape from the bare potentials.  The adiabatic potentials are then altered via a ramp (of variable duration, usually a few ms) of the bias field $B_0$ to a chosen value, with the ramp duration  chosen to prevent vibrational excitation in the dressed lattice.    The detuning $\delta$ and the coupling $\Omega$ are calibrated through observations of three-level Rabi oscillation of the dipole-trapped BEC; under the Hamiltonian of Eq.~2, full transfer from $m_F=-1$ to $m_F=+1$ occurs at a time given by $T = \pi\sqrt{2}/\Omega$, with typical values of $T=1.8(1)$  $\mu$s for $\Omega=2\pi\times$400(20)  kHz near a resonance of 3.85 MHz (corresponding to a magnetic field of 5.48 G).
At the termination of the magnetic-field sweep we either terminate the experiment or hold the field to study the lifetime of the dressed lattice; we then acquire data reflecting the trapped sample's momentum distribution through rapid (sub-$\mu$s) switch-off of the lattice beams, rf field, and background dipole trap, followed by 10--20 ms of ballistic expansion and a typical resonant absorption imaging process, with optional accompanying Stern-Gerlach gradients during time-of-flight to provide spin-projection information.   

\begin{figure}[t!]
\centering
  \includegraphics[width=\columnwidth]{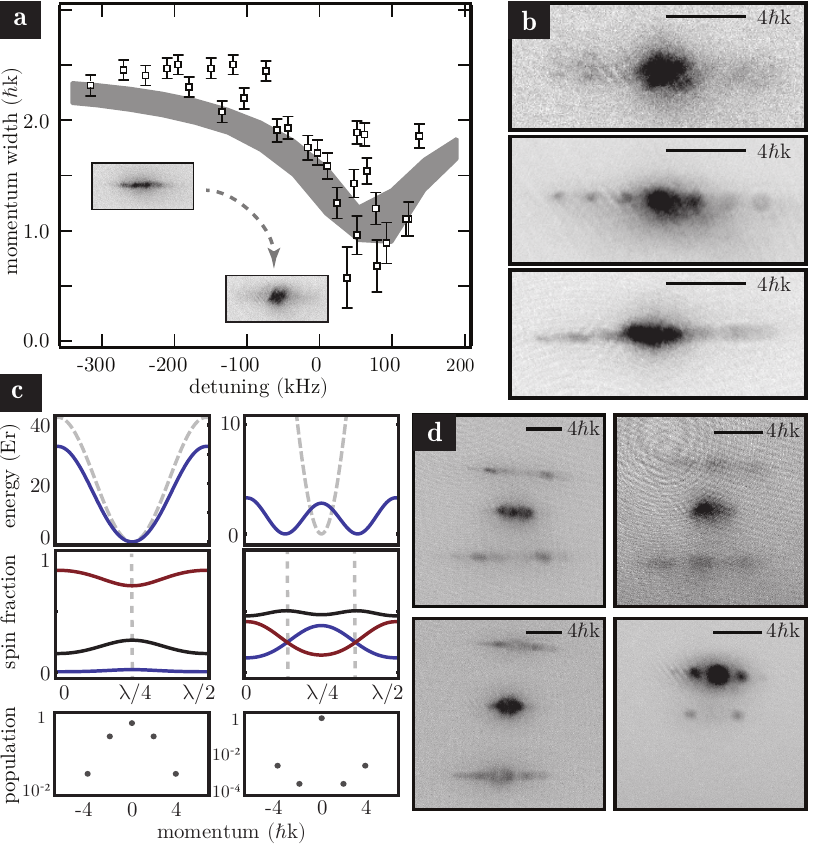}
      \caption{{\bf (a)} Observations of the width (Gaussian $1/e$ radius) of the lattice-trapped atomic momentum distribution as a function of the dressing field detuning $\delta$ for a bare lattice of depth $U_0= 44 E_r$ and coupling strength $\Omega/2\pi = 400$ kHz.     Shaded area represents the predicted equivalent widths from a 1D band-structure calculation accounting for 1$\sigma$ uncertainties in $U_0$, $\Omega$, and $\delta$.    Error bars represent a combination of fit uncertainty and uncertainty in the finite-size correction.   Inset: typical observed momentum distributions for the bare lattice and the maximally-altered lattice.    
 {\bf (b)} Typical examples of the lattice momentum distribution near $\delta_0$.     The visual character of a given iteration of the experiment at  $\delta_0$  depends on fluctuations of $U_0$, $\Omega$, and $\delta$; the dephased examples (top and bottom) were the most repeatable.
  {\bf (c)}  For the data in (a) at $\delta/2\pi=-300$ kHz (left) and $\delta/2\pi=+80$ kHz, near $\delta_0$ (right): top, blue: calculated dressed lattice (uppermost adiabatic potential), dashed: bare lattice.    Middle: the bare spin distribution of the adiabatic eigenstate; $m_F=-1$ (red), $m_F=0$ (black), $m_F=+1$ (blue).    The dashed lines indicate the lattice sites of the adiabatic potential.  Bottom: momentum-component weights of the associated lattice shape, as determined by a 1D band-structure calculation. {\bf (d)}  Stern-Gerlach separation of the dressed lattice; all images near $\delta_0$ except at lower right, taken at $\delta/2\pi = -300$ kHz.    All images in (d) progress from top to bottom as $m_F$: $-$1, 0, +1.
}
\label{momentum}
\end{figure}

Figure~\ref{momentum}~summarizes our observations of the dressed lattice as the final value of $B_0$ is varied, corresponding to alteration of the bare $\lambda/2$-periodic potentials to a regime where significant $\lambda/4$ periodicity is expected (namely, correspondingly increased weight in momentum space at $\pm 4\hbar k$), concomitant with significant reduction in overall lattice depth and associated reduction in width of the momentum-space envelope.   The data in Fig.~\ref{momentum} were taken at a bare lattice depth of $44 E_r$, where $E_r$ is the single-photon recoil energy $\hbar^2 k^2/2 m = h \times 3678$ Hz.    We fit the observed momentum distributions to Gaussian profiles, which yield increasingly poorer fits as the dressing sweep approaches the critical detuning $\delta_0\simeq U_0/2$ but nevertheless provide some information about the system.   We use a 1D band structure calculation informed by the dressing parameters to calculate the expected widths; the small deviation of data from our noninteracting theory is likely driven by number fluctuation, associated uncertainty in the time-of-flight/finite-size correction, and by inaccuracies in the simple fitting function.   As the dressing rf approaches resonance the potential becomes increasingly distorted, and near  $\delta_0$ we see a minimum in the fit width of the momentum profile, but more importantly a significant deviation from a Gaussian profile. We see either a strong central peak with either i) a faint  background at $\pm 4\hbar k$ (Fig.~2(b) top), ii) clearly resolved momentum orders with magnitudes suggestive of a highly distorted lattice  (Fig.~2(b) middle), or iii) a dephased envelope (Fig.~2(b) bottom) very different in character from the typical Gaussian momentum profile of a $\lambda/2$ lattice, as suggested by the 1D band structure calculation of momentum orders shown in Fig.~2(c) at bottom.   The particular appearance of the dressed $\lambda/4$ cloud in a given iteration of the experiment likely depends on fluctuations of $U_0$ and number, as well as a possible dependence on hold time and loading time.  The dephased example was the most repeatable.  For a $\lambda/4$ dressed lattice of calculated 3--4 $E_r$ depth we expect to see faint components at $\pm 4\hbar k$ if the system has maintained some phase coherence, otherwise an envelope including greater weight at higher momenta than would be expected for such a lattice.    Stern-Gerlach separation (with a  $\sim$10 G/cm gradient) can shed some light on the nature of the dressed lattice by projecting the adiabatic eigenstate onto the bare spin states as the lattice is shut off.   The spin superpositions of Fig.~2(c) suggest that at $\delta_0$ the lattice atoms should have equal weight of $m_F=\pm 1$ and 50\% weight of $m_F=0$.     The images in Fig.~2(d) confirm this, and also reveal the clear signature of significant weight at $\pm 4\hbar k$.     A dressed-state Bloch-band model (not shown) confirms the spin-weights of the dressed lattice at a variety of bare depths, coupling strengths, and detunings.


\begin{figure}[t!]
\centering
  \includegraphics[width=\columnwidth]{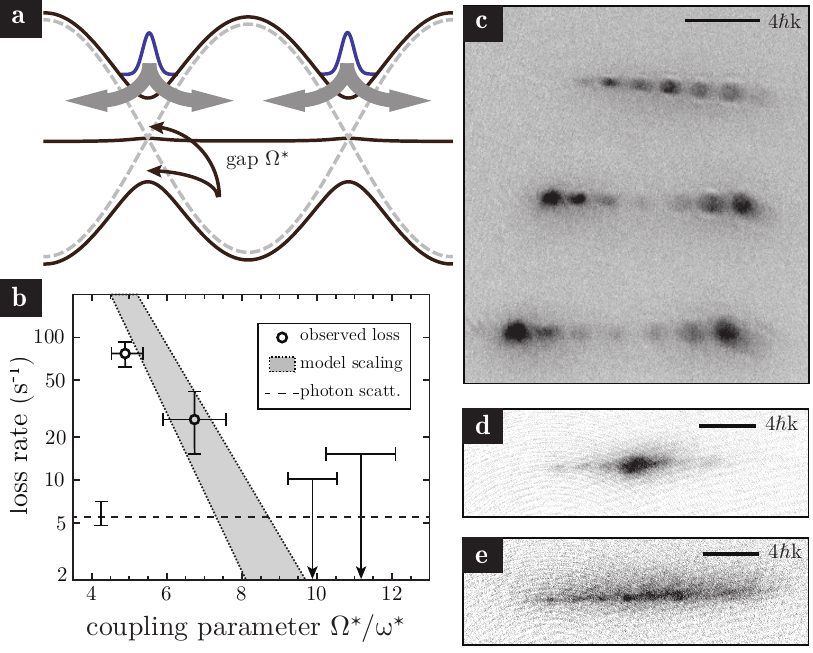}
      \caption{{\bf (a)} Schematic of the loss mechanism from the uppermost adiabatic potential.    The gap $\Omega^*$ (proportional to $\Omega$) exponentially suppresses nonadiabatic transitions at the avoided crossing; increasing this gap comes at a cost of reducing dressed lattice depth.    Use of the lower adiabatic potential, while stable against loss, is unproductive (see text).   {\bf (b)}~Loss data as a function of the dimensionless dressing parameter $\Omega^*/\omega^*$, where $\omega^*$ is the oscillation frequency around the adiabatic-potential minima.    Note logarithmic scale; the rightmost two points represent loss below a given upper limit resulting from noise and finite hold time.  The shaded area represents the scaling of Eq.~3 for calculated values of the coupling parameter, with the uncertainty stemming from the model's range of $\alpha$ and uncertainty in $\delta$.    The dashed line represents the expected photon-scattering rate from the lattice laser.  {\bf (c)}   The loss process is most easily visualized using Stern-Gerlach separation immediately following a dressing attempt; here we applied a very weak ($\Omega/2\pi\simeq$ 30 kHz) dressing field.     The clear difference between successful and lossy dressing can also be seen in the combined momentum distributions following a dressing attempt: {\bf (d)} a typical near-$\delta_0$ iteration of the long-lived strongly-dressed lattice, in contrast to {\bf (e)} the short-lived high momentum spread of a weakly-dressed lattice.}      
\label{loss}
\end{figure}


Several issues regarding adiabaticity are present in this system.   Spin-following during the period of the sweep $\delta(t)$ where the adiabatic potentials are roughly identical to the bare potentials is easily maintained with a typical rate of $\dot{\delta}=$250 kHz/ms,  satisfying $\dot{\delta}\ll \omega_{-1,0}^2$.     More dangerous is the potential for too-rapid distortion of the dressed lattice leading to vibrational excitation.   While the usual criterion~\cite{PhysRevLett.76.4508} suggests that  sub-ms ramps are safe, since the lattice is initially dephased (and thus filling the first Brillouin zone) we use a conservative ramp duration (2--4 ms).      Most importantly, the degree to which the atoms respect the adiabatic eigenstates determines the lifetime of the dressed lattices; in the case of weak coupling, the adiabatic potentials  are not an accurate predictor of the atomic dynamics.     Nonadiabatic transitions should show up as loss under our experimental protocol through the transition to high-lying momentum states of the lower adiabatic potentials; intuitively this can be viewed as a Landau-Zener problem at an avoided crossing.        To measure this we simply load the lattice under a given set of dressing parameters and hold it for a variable duration.    These data are summarized in Fig.~3, where we compare the number of atoms remaining in the uppermost adiabatic potential over a variable hold time, while varying the dressing configuration.  Crucially, for dressing parameters $\Omega^*/\omega^* \gtrsim 10$ we observe an upper limit on nonadiabatic loss, implying stability on timescales $\gtrsim$ 30$+$ ms.  This approaches the expected photon-scattering lifetime of $\sim$100 ms. The nature of the loss is most clear in Fig.~3(c,e) where a loading sequence was performed with very weak dressing field; nonadiabatic transitions are dominant, and result in high-lying momentum states of the lower two adiabatic potentials.     

The details of this loss mechanism have been treated  theoretically in Ref.~\cite{Yi:2008hy}, predicting a loss rate behaving exponentially in the avoided-crossing gap $\Omega^*$ as well as in the dressed-lattice trap frequency $\omega^*$:
\begin{equation}
\Gamma =\Gamma_0 \exp{(-\alpha\Omega^*/\omega^*)}
\end{equation}
where $\Gamma_0$ is roughly the Landau-Zener attempt frequency $\omega^*/2\pi$ (which depends weakly on $\Omega$), with $\alpha = 1.1(1)$.     The observed strong increase in lifetime over the typical data of Ref.~\cite{Lundblad:2008kb} stems from a tripling of the gap $\Omega^*$ which permits a new regime of dressed lattice structure.    One possible route to avoid these losses altogether would be to use the lowest adiabatic potentials~\cite{2008PhRvA..78e1602S}, and sweep the detuning downward rather than upward in order to distort the lattice; in this case the degree to which the system was adiabatic would be difficult to observe, as the atoms transitioning to the bare lattice would not be lost, but rather occupy the same space while obeying a different band structure.   In the limit of strong enough dressing such that the experimenter was sure that losses were minimal this would perhaps be an attractive technique. 

While the dressed lattices we observe are still relatively weak, they are a clear improvement over the typical dressed depths in Ref.~\cite{Lundblad:2008kb} which were of order a single recoil.  To push the depths of the dressed lattice to tens of recoils will require the deepening of the bare lattice, which will initially increase losses; however, the scaling of Eq.~3 is promising in that losses can be suppressed simply by increasing $\Omega$ and $U_0$.      Of course, regardless of strong dressing, lifetimes beyond 100 ms  will be prevented in this system by photon scattering from the lattice beams.      We seek to extend the band-structure-alteration capabilities of this technique, anticipate using this new regime of strong $\Omega$ in two- and three-dimensional lattices to explore Bose-Hubbard physics on a reduced-periodicity lattice, create toroidal single-site wavefunctions, and study the possibility of using multiple dressing frequencies~\cite{Yi:2008hy,Morgan:2014vd} to generate even finer subdivision of the lattice.  


\begin{acknowledgments}
We thank Nathaniel Funk, Janith Rupasinghe, Marc Tollin, and Joanna Moody for early work on the apparatus.   We also thank Bates College for generous support.   This work was partially supported by AFOSR/DEPSCoR and NSF MRI.  
\end{acknowledgments}

\bibliographystyle{apsrevNOURL}
\bibliography{/Users/nlundbla/Desktop/latex-stuff/master3}

\end{document}